\documentclass[sigconf, nonacm, screen]{acmart}

\usepackage[noend]{algpseudocode}
\usepackage{float}
\usepackage{subcaption}
\usepackage{tabularx}
\usepackage{booktabs}
\usepackage{algorithm2e}
\usepackage{multirow}
\usepackage{booktabs}
\usepackage{tablefootnote}
\usepackage{threeparttable}
\usepackage{mdframed}
\usepackage{minted}
\usepackage{placeins}
\usepackage{threeparttable}

\usepackage{amsmath,amssymb}
\usepackage{pifont}
\usepackage{xcolor}
\usepackage{listings}

\definecolor{codegreen}{rgb}{0,0.6,0}
\definecolor{codegray}{rgb}{0.5,0.5,0.5}

\definecolor{backcolour}{RGB}{245,248,250}
\definecolor{emph}{RGB}{166,88,53}
\definecolor{nightblue}{RGB}{9,49,105}
\definecolor{keywords}{RGB}{207,33,46}
\definecolor{lightpurple}{RGB}{130,81,223}

\lstdefinestyle{mystyle}{
    backgroundcolor=\color{backcolour},   
    commentstyle=\color{codegreen},
    keywordstyle=\color{keywords},
    stringstyle=\color{nightblue},
    basicstyle=\ttfamily\footnotesize,
    breakatwhitespace=false,         
    breaklines=true,                 
    captionpos=b,                    
    keepspaces=true,                 
    numberstyle=\small\color{codegray},
    numbers=left,                    
    numbersep=5pt,   
    xleftmargin=0.2cm,
    aboveskip=0.2cm,
    belowskip=0.1cm,
    showspaces=false,                
    showstringspaces=false,
    showtabs=false,                  
    tabsize=2,
    frame=shadowbox,
    emph={},
    emphstyle={\color{lightpurple}},
}

\definecolor{pastelyellow}{RGB}{255, 255, 230}
\definecolor{lightorange}{RGB}{255, 223, 186}
\definecolor{orange}{RGB}{255, 140, 0}

\AtBeginDocument{%
  \providecommand\BibTeX{{%
    \normalfont B\kern-0.5em{\scshape i\kern-0.25em b}\kern-0.8em\TeX}}}

\acmPrice{15.00}
\acmISBN{978-1-4503-XXXX-X/18/06}

\begin{document}

\title{What Happens When the Model Eats the Stack? Rethinking the Research Agenda for Data Agents to Withstand the Bitter Lesson}


\author{
    Liana Patel\textsuperscript{2}\,
    Siddharth Jha\textsuperscript{2}\,
    Negar Arabzadeh\textsuperscript{1}\,
    Carlos Guestrin\textsuperscript{2}\,
    Ion Stoica\textsuperscript{1}\,
    Matei Zaharia\textsuperscript{1}\\
    \normalsize \textsuperscript{1}UC Berkeley \quad
    \textsuperscript{2}Stanford University \\
}

\renewcommand{\shortauthors}{}

\newif\ifcomments
\commentstrue
\ifcomments
    \providecommand{\liana}[1]{{\color{blue}{/* liana: #1 */}}}
    \providecommand{\sid}[1]{{\color{purple}{/* sid: #1 */}}}
\else
    \providecommand{\liana}[1]{}
    \providecommand{\sid}[1]{}
\fi

\newif\ifgap
\gaptrue
\ifgap
    \providecommand{\gap}[1]{{\color{purple}{/* gap: #1 */}}}
    \providecommand{\sid}[1]{{\color{red}{/* Sid: #1 */}}}
\else
    \providecommand{\gap}[1]{}
\fi

\newif\ifrev
\ifrev
    \providecommand{\rev}[1]{{\color{blue}{#1}}}

\else
    \providecommand{\rev}[1]{{\color{black}{#1}}}

\fi



\newcommand{\heading}[1] {{\hfill\break\noindent{\textbf{\emph{#1}}} }}
\newcommand{\comma}[1] {{\textit{,\space\space}{#1} }}
\newcommand{\tagbench}[1]{{TAG-bench\xspace}}

\begin{abstract}
The bitter lesson poses an existential question for the data systems community, whereby large language models (LLMs) trained end-to-end are rapidly internalizing new capabilities that previously required carefully engineered data agents.
Guided by empirical insights, we argue that as models continue to improve, many proposed system layers designed to compensate for model limitations on a given task will increasingly be subsumed by the model itself.
We instead identify enduring research opportunities, which lie in supporting data agents \emph{across many queries} with curated contextual information about the data environment, which we call \emph{persistent semantic context}.
We find that these context layers demonstrate strong promise for improving data agent performance, but they also raise significant system challenges.
Thus, a key requirement for future data systems will lie in natively serving persistent semantic contexts as a first-class abstraction in order to enable capable data agents working over huge, complex knowledge corpora.
Towards this vision, we outline exciting new research opportunities, including designing efficient context data structures, storage methods, compression techniques, and semantic consistency protocols, to ensure integrity and correctness of the stored contextual knowledge.
\end{abstract}






\maketitle

\section{Introduction}
\label{sec:intro}
Driven by increasingly powerful frontier large language models (LLMs), agentic capabilities are undergoing a dramatic transformation that rapidly shifts the landscape for data agents and promises to enable users' natural language questions over their data. 
Once simple text generators, LLMs have evolved over the past several years to autonomously execute complex, long-horizon tasks by reasoning, invoking tools, executing code, and forking subagents when paired with a terminal, file system and execution loop.

This trajectory poses an existential question for the data systems community, one crystallized by Sutton's \emph{bitter lesson}~\cite{sutton2019bitter}: general methods that scale computation (e.g., end-to-end model training) ultimately displace hand-engineered domain knowledge. 
Consequently, we ask the question, \emph{as models continue to improve, threatening to eat the stack with internalized capabilities, what systems problems will endure and become more crucial for enabling future data agents?}

To ground this question empirically, we study the evolution of agentic capabilities for data processing over the past two years. We measure the performance of both state-of-the-art, human-designed data agents and general coding agents on recent data agent benchmarks using frontier models from 2025 to 2026. Our experimental analysis reveals surprising insights, which counter conventional wisdom on agent-first system design. 
First, we find that as model capabilities improve, general coding agents substantially outperform human-designed data agents, achieving higher accuracy and obviating the need for these pipelines that impose human priors through rigid scaffolding and task decomposition. Moreover, newer models are rapidly improving the \emph{efficiency} of general coding agents, drastically reducing the average number of turns and tokens required per query.  
This trend directly challenges the premise of recent work~\cite{liu2025overlords}, which argues that future data systems will be bottlenecked by \emph{agentic speculation}, a workload characterized by sheer scale and inefficiency of agent-issued queries, which data systems will need to optimize for. We instead see that newer models natively improve efficiency on a given task, solving each one with fewer actions and less frequent failures, on average, over successive model generations.

Interestingly, we find that significant challenges remain in supporting data agents \emph{across many queries} with rich contextual information about the data environment. 
Even the strongest agents must repeatedly re-acquire context about their environment, incurring significant relative costs to rediscover schema, join paths, and data semantics that prior queries already surfaced. 
These inefficiencies that persist are not failures of \emph{reasoning}, which models are rapidly internalizing, but rather they are failures of \emph{knowledge about a particular data environment}, which is external to the model, changes as data changes, and may be wastefully re-derived.

As a result, an increasingly important requirement for more capable data agents will lie in non-parametric contextual information, i.e., explicit knowledge of the working environment that the model does not acquire during training as parametric knowledge. We refer to this contextual knowledge as \emph{persistent semantic context}. This state will be constructed offline to provide a language-based representation of knowledge about the data environment, and its associated costs can be amortized across many user queries. We demonstrate the promise of semantic context, finding that even simple, agent-authored contexts (e.g., files that the agent writes offline given historical traces and access to the data environment) can significantly improve accuracy and reduce per-task exploration, a finding likewise substantiated in recent work~\cite{agentsm2026, lin2025sleeptime, agrawal2025gepa}. 
However, we also find that persistent semantic contexts introduce significant system overheads, creating important new challenges, which our community is well-positioned to address.

\begin{figure*}[!th]
  \centering

  \includegraphics[width=0.62\textwidth]{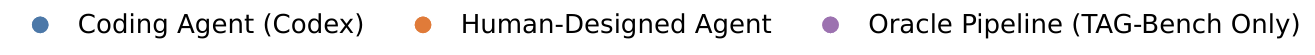}
  \vspace{-0.5em}

  \begin{subfigure}[t]{0.49\textwidth}
    \centering
    \includegraphics[width=\linewidth]{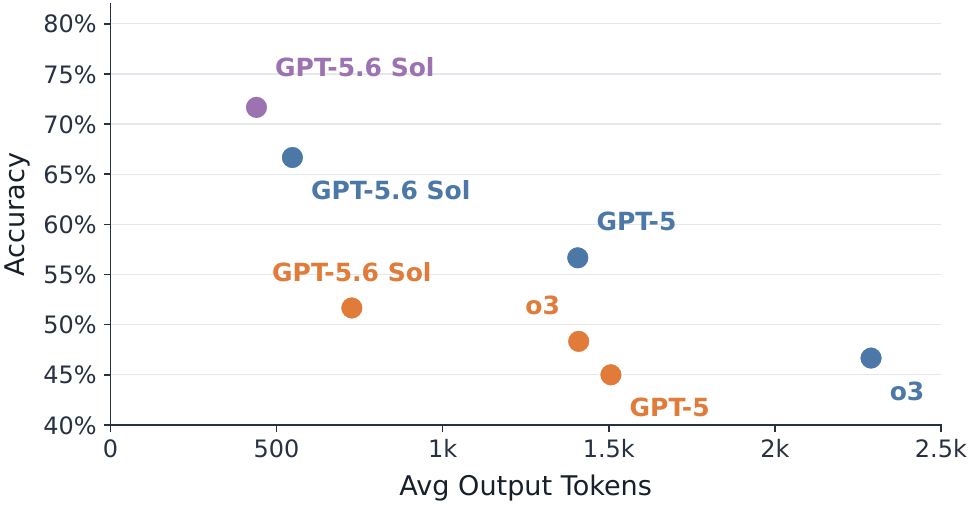}
    \caption{TAG-Bench}
    \label{fig:tag-evolution}
  \end{subfigure}
  \hfill
  \begin{subfigure}[t]{0.49\textwidth}
    \centering
    \includegraphics[width=\linewidth]{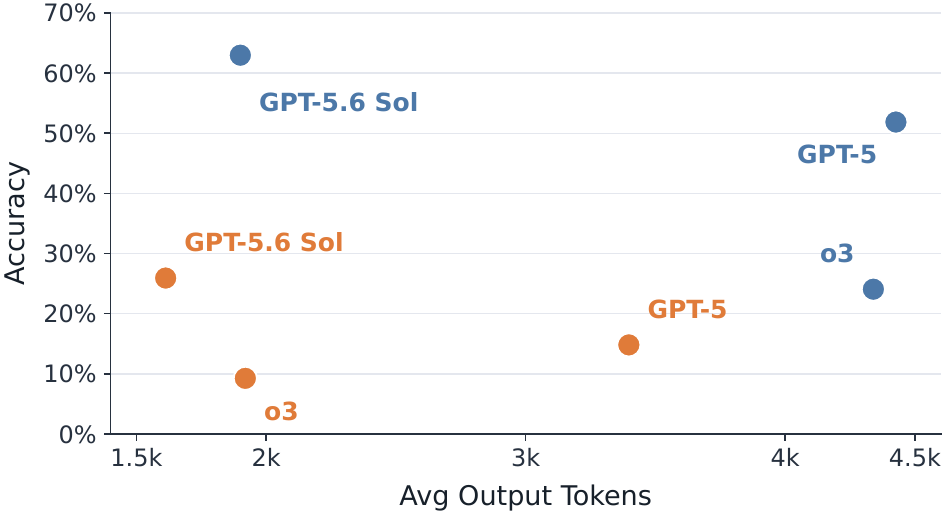}
    \caption{DAB}
    \label{fig:dab-evolution}
  \end{subfigure}

  \caption{\textbf{Performance of general coding agents (blue) and human-designed
  data agents (orange) with successive model versions, including o3 (released early 2025), GPT-5 (released mid-2025), and GPT-5.6 Sol (released mid-2026) on two data-agent benchmarks, TAG-Bench and DAB.}}
  \label{fig:evolution}
\end{figure*}

In order to serve more capable data agents, future data systems will need to natively serve persistent semantic contexts as a first-class abstraction. Towards this vision, we outline new research opportunities, including designing new semantic consistency protocols to ensure the integrity and correctness of stored contextual knowledge, as well as designing efficient physical layouts, data structures and compression techniques for this layer. 
In summary, we make the following contributions:
\begin{itemize}
    \item We study the evolving capabilities of general coding agents and human-designed data agents over the past two years, measuring their task performance, efficiency, and failure modes on recent data agent benchmarks.
    \item We demonstrate that while increasingly capable models are improving performance and efficiency for individual tasks, a significant enduring challenge lies in supporting data agents across many queries with persistent semantic context, which provides rich contextual information about the data environment.
    \item Based on these insights, we outline new research opportunities that target a key requirement of future data systems: native support of persistent semantic contexts for agents.
\end{itemize}

\section{Understanding the Evolution of Agent Capabilities and Failures}
\label{sec:eval}

In this section, we study how agentic capabilities for data processing have evolved over the past two years, comparing the performance of general coding agents and state-of-the-art, human-designed data agents. We use two recent data agent benchmarks: TAG-Bench~\cite{biswal2024tag}, which consists of queries that require combining exact computation, semantic reasoning, and world knowledge over relational databases, and Data-Agent Benchmark (DAB)~\cite{dab2025}, which evaluates multi-step analytical tasks over fragmented enterprise data. We evaluate three model families spanning 2025 to 2026: o3~\cite{openai_o3_2025}, released in early 2025; GPT-5~\cite{openai_gpt5_2025}, released in mid-2025; and GPT-5.6 Sol~\cite{openai_gpt56_sol_2026}, released in mid-2026. We evaluate each model using the Codex coding agent harness and additionally compare against state-of-the-art human-designed data agents. On TAG, we evaluate the human-designed data agent, Agentar-Scale-SQL~\cite{wang2025agentarscalesqladvancingtexttosqlorchestrated}, a state-of-the-art system on the BIRD leaderboard~\cite{bird} whose full inference code is not publicly available. We use its published SQL-generation prompt and add a custom execution-guided refinement step based on the procedure described in the paper. On DAB, which involves multi-turn tasks that Agentar-SQL is not designed for, we evaluate DeepEye~\cite{Li_2026}, the next-best-performing open-source system on the BIRD-SQL leaderboard that can perform multi-turn tasks. DeepEye is a multi-agent system that first generates a data-analysis workflow, then executes and repairs it as needed. For each baseline, we sweep all model reasoning levels and report results with low reasoning for simplicity, finding consistent trends at other levels.

\heading{With more capable models, general coding agents substantially outperform human-designed data agents.}
Figure~\ref{fig:evolution} reports the performance of general coding agents and state-of-the-art human-designed data agents on TAG-Bench and DAB. Notably, we see that with \texttt{GPT-5.6 Sol}, the general coding agent achieves significantly higher accuracy on both benchmarks than the human-designed data agent using the same model. Interestingly, with the weaker \texttt{o3} model, on TAG-Bench, the human-designed \texttt{Agentar-Scale-SQL} data agent achieves higher accuracy and better token efficiency than the o3 coding agent. However, over successive model generations, this trend reverses, with the coding agent achieving both better accuracy and token efficiency than the human-designed data agent with the same models. These results indicate that as more capable models internalize capabilities such as planning, code generation, debugging, and validation, their performance and generalization will likely surpass that of data agents with human-imposed priors and fixed designs.

\heading{General coding agents are rapidly improving both accuracy and token efficiency.}
Figure~\ref{fig:evolution} shows that on TAG-Bench, the Coding Agent improves both accuracy and token efficiency with successive model versions. In fact, the \texttt{GPT-5.6 Sol Coding Agent} achieves accuracy comparable to that of the Oracle baseline, which consists of expert-written queries run with the LOTUS runtime~\cite{patel_lotus_2024}. Similarly, on DAB, the \texttt{GPT-5.6 Coding Agent} significantly outperforms prior agents, gaining over 35 percentage points over the \texttt{o3 Coding Agent} while improving token efficiency by over $2\times$. These gains come from a \emph{general} agent harness with no task- or data-specific engineering.

\begin{figure}[t]
  \centering
    \includegraphics[width=\linewidth]{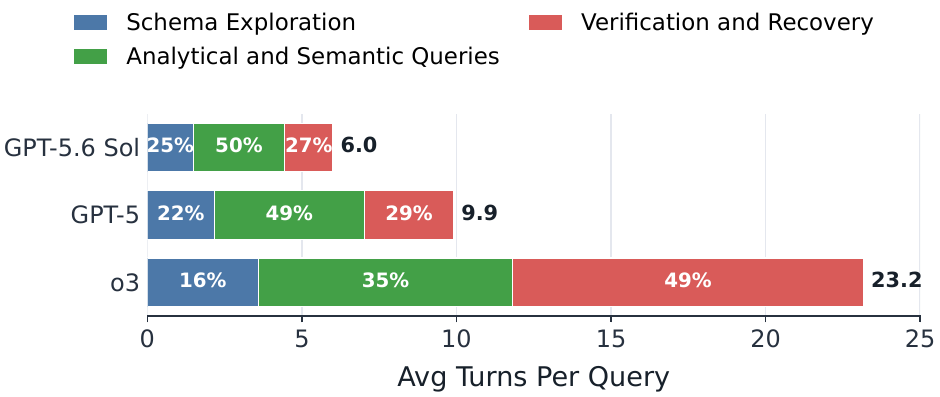}
  \caption{Turns per query for Coding Agents across model generations on DAB.}
  \label{fig:composition}
\end{figure}

\heading{Newer models drastically reduce the number of turns per query, improving token efficiency.}
We study the efficiency of the \texttt{Coding Agents} by measuring the number of turns per query on DAB, decomposed by action types such as schema exploration, analytical computation, and verification.
Figure~\ref{fig:composition} demonstrates a sharp decline in the number of turns per query required by coding agents with successive model generations. 
Notably, the \texttt{GPT-5.6 Sol Coding Agent} reduces the number of turns per task by about $4\times$ compared to the \texttt{o3 Coding Agent}. This efficiency gain is driven by significantly fewer turns spent on analytical and semantic querying, as well as fewer verification or recovery steps.
Interestingly, this trend counters the premise suggested by prior work~\cite{liu2025overlords}, which presumes that agentic workloads will create a high volume of speculative, inefficient queries for data systems. Instead, we should expect models to continue becoming more efficient, formulating correct answers with fewer turns per query.

\heading{The relative cost of schema exploration has increased over successive model generations.}
We also see from Figure~\ref{fig:composition} that while more capable models spend fewer turns overall, steps spent on schema exploration or data profiling remain a substantial relative cost for the \texttt{GPT-5.6 Sol Coding Agent}. As reasoning, planning, and debugging capabilities improve, a growing fraction of agent effort is devoted to acquiring information that the model cannot know a priori: what data exists, what it means, and which sources are relevant to the task at hand. We expect that as models become more capable, a substantial cost will shift from iteratively formulating queries and validating results to understanding the data environment itself.

\begin{figure}[!t]
  \centering    \includegraphics[width=\linewidth]{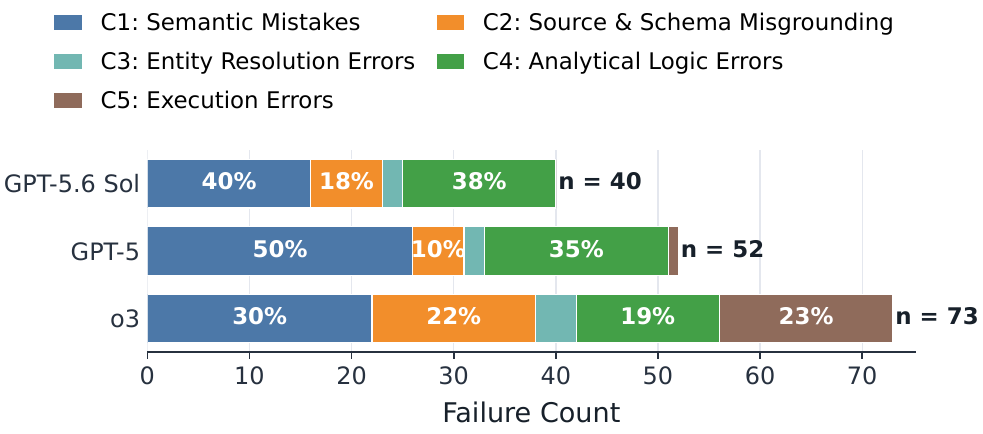}
  \caption{\textbf{Failures by root-cause cluster for Coding Agents across model generations on TAG and DAB.}}
  \label{fig:failures}
\end{figure}

\heading{Existing challenges lie in environmental and contextual understanding.}
To understand trends in failure modes and remaining gaps, we analyze the trajectory of each failed \texttt{Coding Agent} query on TAG and DAB. We use \texttt{GPT-5.6 Sol} to assign each trace to a root-cause cluster, following a taxonomy of five key failure clusters described in Table~\ref{tab:failure-clusters}. Figure~\ref{fig:failures} shows the distribution of failures for both benchmarks. We find that execution errors due to incorrect tool use or execution decline as model capability increases. Meanwhile, the remaining errors for GPT-5.6~Sol are dominated by \emph{failures of environmental knowledge}. Specifically, over $60\%$ of failures arise from semantic misinterpretation, including misunderstanding task meaning or relying on invalid proxies (C1), using the wrong data sources, tables, fields, or metrics (C2), or incorrectly using entities, join keys, and identifiers.

\begin{figure*}[!t]
  \centering
  \begin{subfigure}[t]{0.49\textwidth}
    \centering
    \includegraphics[width=\linewidth]{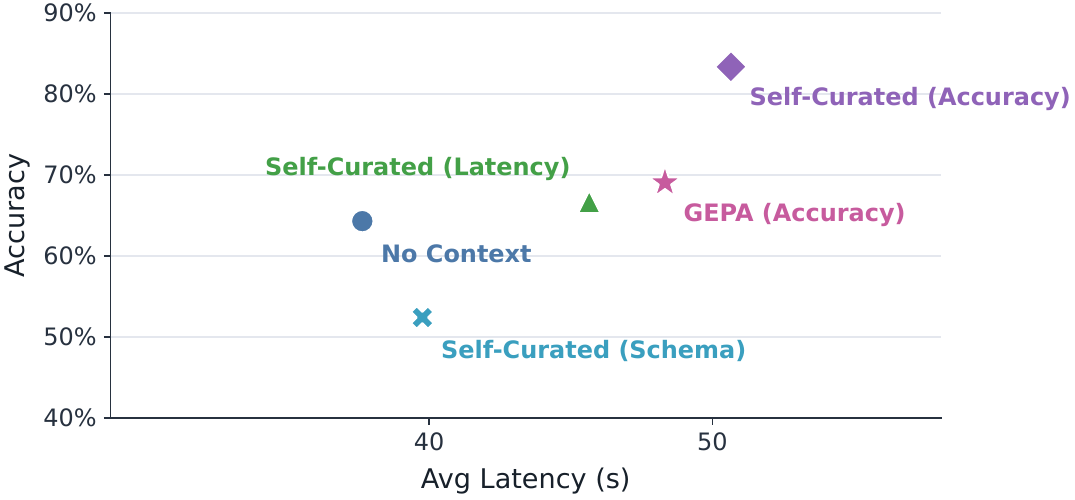}
    \caption{Coding Agent Performance with Different Persistent Contexts}
    \label{fig:ctx-res}
  \end{subfigure}
  \hfill
  \begin{subfigure}[t]{0.49\textwidth}
    \centering
    \includegraphics[width=\linewidth]{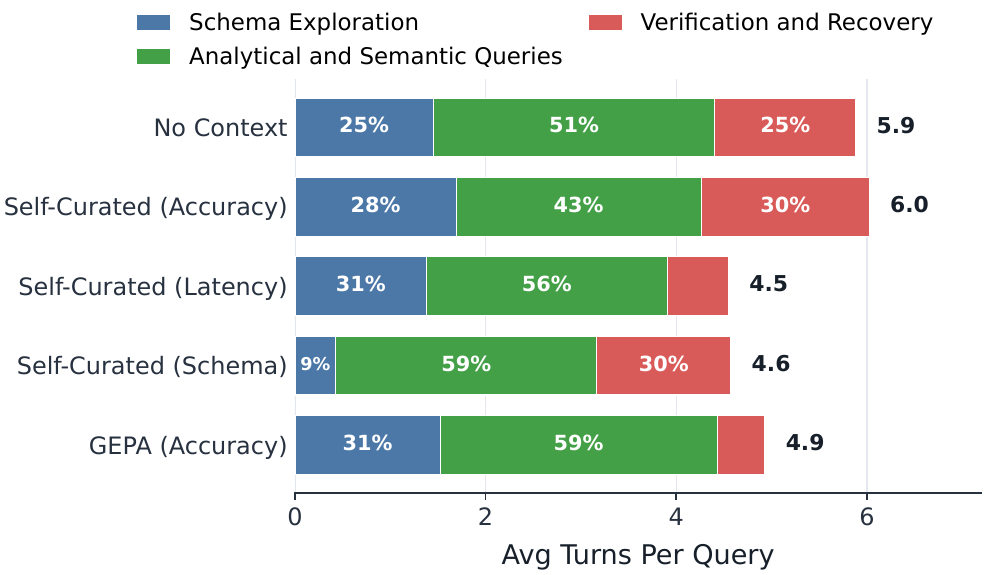}
    \caption{Turns per query.}
    \label{fig:ctx-composition}
  \end{subfigure}
  \caption{Performance of the GPT-5.6 Sol Coding Agent with additional Self-Curated or GEPA-generated context provided during task execution on held-out DAB tasks. The context is constructed offline using GPT-5.6 Sol and 12 sample trajectories to target different metrics (Latency, Accuracy, and Schema Knowledge), then appended to the model prompt during evaluation.}
  \label{fig:ctx-eng}
\end{figure*}

\begin{table}[t]
\centering
\footnotesize
\caption{Unified taxonomy of root-cause failure clusters.}
\begin{tabular}{@{}l p{0.82\columnwidth}@{}}
\toprule
 & Failure Cluster \\
\midrule
C1 & Misinterpreted task meaning or reliance on an invalid proxy \\
C2 & Wrong data source, table, field, or metric \\
C3 & Entity, join-key, or identifier mismatch \\
C4 & Incorrect filtering, aggregation, ranking, or output logic \\
C5 & Runtime, tooling, or command-execution failure \\
\bottomrule
\end{tabular}
\label{tab:failure-clusters}
\vspace{-.3cm}
\end{table}

\heading{Persistent semantic context can improve performance and amortize per-task exploration, but introduces significant system overheads.} To study the performance implications of leveraging richer contextual information, we evaluate the \texttt{GPT-5.6 Coding Agent}, providing it with additional persistent context generated offline using different techniques. Specifically, we generate \texttt{Self-Curated} contexts by prompting the coding agent to author its own context, targeting various performance goals, including reducing latency (\texttt{Latency}), improving accuracy (\texttt{Accuracy}), or capturing schema knowledge to reduce online exploration (\texttt{Schema}). We also evaluate \texttt{GEPA}~\cite{agrawal2025gepa}, targeting improved accuracy. We run each context-generation baseline with \texttt{GPT-5.6 Sol}, providing it with access to 12 sample queries and all databases in DAB. During evaluation, we run the \texttt{GPT-5.6 Sol Coding Agent} on all remaining held-out DAB tasks, appending the generated context to the user prompt. We compare against the \texttt{No Context} baseline, which appends no persistent context to the model input but allows the model to explore its environment per query, as in the previous experiments.

Figure~\ref{fig:ctx-eng} demonstrates that agent-generated contexts can effectively improve performance along various target metrics. As Figure~\ref{fig:ctx-res} shows, the \texttt{Self-Curated (Accuracy)} context produces a significant accuracy gain of 19 percentage points over the \texttt{No Context} Coding Agent.
Interestingly, the \texttt{Self-Curated (Latency)} context reduces the average number of agent turns per query and latency compared to the Accuracy-based contexts, but still increases latency over the \texttt{No Context} baseline, as seen in Figure~\ref{fig:ctx-composition}. 
On the other hand, the \texttt{Self-Curated (Schema)} context effectively captures knowledge about the data environment, allowing the agent to spend significantly fewer turns per query on schema exploration during evaluation, as Figure~\ref{fig:ctx-composition} demonstrates. However, the overall Coding Agent accuracy drops modestly, as shown in Figure~\ref{fig:ctx-res}, possibly due to the agent's over-reliance on the curated context for some tasks and overfitting to the small 12-trajectory sample.

\begin{table}[t]
\vspace{-.2cm}
\centering
\caption{Build time, build cost, and memory overhead of techniques for context construction.}
\small
\begin{tabular}{l rrr}
\toprule
 & Build Time & Build Cost & Context Size (KB)\\
\midrule
Self-Curated (Accuracy) & 164.9s & \$1.09 & 2.52 \\
Self-Curated (Latency)  & 154.9s & \$1.13 & 2.25 \\
Self-Curated (Schema) & 3413.6s & \$9.60 & 162.14 \\
GEPA (Accuracy) & 1359.8s & \$12.16 & 7.55 \\
\bottomrule
\end{tabular}
\vspace{-.4cm}
\label{tab:prompt-gen}
\end{table}

Turning to Table~\ref{tab:prompt-gen}, we see that constructing and storing the persistent context can be costly, with substantial build times, build costs, and memory overheads that raise new system challenges. Notably, our setting constructs contexts in a small-scale environment, with only 12 historical traces and 12 datasets, whereas realistic data agent environments will contain terabytes of data involving thousands of tables and long-horizon agent traces. Even in our small-scale experimental setting, constructing the contexts required hundreds to thousands of seconds. The \texttt{Self-Curated (Schema)} context exhibited the highest costs, with a build time of over $3,400$ seconds, a cost of $\$9.60$, and a size of $162$ KB. These overheads result from the agent performing an expansive offline search over the datasets and storing a large amount of knowledge about the data environment to reduce online exploration. We expect the system costs associated with building, storing, and maintaining persistent contexts to scale with the size and complexity of the data store, as well as with the scope and horizon of agent tasks.

\section{Towards a Bitter-Lesson-Proof Research Agenda}
\label{sec:agenda}
As models continue to improve, many capabilities that previously required carefully engineered agent pipelines—including planning, code generation, tool use, debugging, and iterative validation—are increasingly provided directly by the underlying model. Consequently, system layers whose primary purpose is to compensate for weaknesses in these capabilities are likely to diminish in importance over time. This trend reflects the Bitter Lesson~\cite{sutton2019bitter}: improvements in general learning systems ultimately outperform increasingly sophisticated hand-engineered algorithms and heuristics.

However, even highly capable models cannot rely solely on their parametric knowledge when interacting with large, complex data environments. To answer user requests accurately, they must leverage organization-specific knowledge about the current data environment. For example, answering a request such as "Write a quarterly business review explaining why European revenue missed forecast, reconcile the discrepancy against Finance's official numbers, and recommend corrective actions" requires identifying authoritative datasets across multiple systems, resolving conflicting metric definitions, understanding organization-specific business conventions, interpreting semi-structured documentation, and validating intermediate results.
The underlying data environments may span terabytes---significantly larger than typical code repositories---and contain dynamic, organization-specific, and undocumented conventions. This makes contextual information crucial for task completion, and the cost of repeatedly re-acquiring it high.

We therefore lay out a research agenda for the database community centered on supporting \emph{\textbf{persistent semantic context}}, an explicit, language-based representation of knowledge about the data environment that is constructed offline, efficiently accessed online, and amortized across many queries.
The value and volume of this layer will continue to increase over time as models become more capable and user tasks become more complex, growing in both scope and task horizon, requiring the agent to retain external information such as canonical metric definitions, authoritative data sources, entity relationships, historical discoveries, organizational conventions, and user-specific preferences.
Moreover, persistent semantic context can be effectively authored by the model itself, indicating that its quality and size will natively scale alongside model capability as better models are able to leverage and maintain more context and more effectively distill learned environmental knowledge into higher-quality persistent context.
Complementing techniques that involve more costly model training and parametric weight updates~\cite{tandon2025endtoendtesttimetraininglong,zweiger2025selfadaptinglanguagemodels,shenfeld2026selfdistillationenablescontinuallearning}, the context layer will need to retain an explicit knowledge state that is easily editable, fresh, reliable, and human-auditable.
Supporting this functionality will require solving substantial new challenges, creating exciting opportunities for the database community.
We outline key directions for the research community, which will require developing new techniques for semantic consistency, to ensure the correctness and integrity of persistent contexts, as well as optimized data structures, construction methods, physical designs, and compression techniques.

\subsection{Maintaining Semantic Consistency}
\label{subsec:agenda-maintenance}
Over time, the data environment will change as new tables are introduced, schemas evolve, documentation is updated, business definitions are rewritten, users provide corrections, statistical shifts occur in the underlying data, and query patterns change. These updates can invalidate the persistent context, leading to incorrect or inefficient agent executions. This raises a substantial new challenge, which we refer to as \textbf{\emph{semantic consistency}}, i.e., ensuring that the natural language or semantic knowledge contained within the context layer is up-to-date and correct. Unlike traditional consistency models, which typically reason about explicit data objects and operations such as reads, writes, and updates, semantic consistency is challenging because context layers contain implicit relationships between natural-language artifacts. A change to one concept, dataset, or workflow may invalidate seemingly unrelated context entries whose dependencies are not explicitly represented, requiring systems to reason about changes in meaning rather than changes in state. Below we outline several challenges and exciting directions for further research.

\heading{Consistency Scope.} Future context layers will likely require mechanisms analogous to namespaces or multi-version data management, allowing multiple semantic views to coexist while ensuring consistency within their intended scope. At the finest granularity, an individual user may maintain a personalized context layer that captures their preferred datasets, prior interactions, and accumulated execution history. At a broader scope, project teams often share common business definitions, recommended datasets, procedural guidance, and examples that should remain synchronized across collaborators. At the enterprise level, updates to governance policies, trusted data products, or canonical business metrics should propagate consistently across all agents operating within the organization. Finally, some context originates outside the enterprise—for example, public documentation, regulatory guidance, or external APIs—introducing a global scope in which organizational context must remain synchronized with external sources. Determining the appropriate scope of a semantic update therefore becomes a fundamental systems problem. When new knowledge is acquired, the system must determine who should observe the update, how it should propagate across overlapping scopes, and when conflicting semantic views should be reconciled or allowed to coexist.

\heading{Consistency Models.} Once the scope of consistency has been established, the next question is what guarantees the system should provide. Classical database systems offer a rich spectrum of consistency models, ranging from strong consistency to eventual consistency, each reflecting a different tradeoff between freshness, availability, and maintenance cost. An analogous design space exists for semantic context. Applications such as regulatory reporting or production dashboards may require strong semantic consistency, ensuring that every agent observes the latest approved context before execution. Conversely, collaborative analytics may tolerate eventual semantic consistency, allowing semantic updates to propagate asynchronously while reducing maintenance overhead. Finally, systems may adopt query-triggered semantic consistency, lazily revalidating portions of the context layer only when they are accessed by an agent. Such an approach amortizes maintenance costs across future workloads and may be particularly attractive for large context layers containing thousands of semantic artifacts. Determining the appropriate semantic consistency model—and understanding its implications for accuracy, latency, scalability, and maintenance cost—remains an open research question.

\heading{Consistency Methods.}
Once inconsistency has been detected, the system must decide how to restore consistency. One approach is incremental semantic maintenance, where the context layer is decomposed into fine-grained semantic artifacts that are selectively re-validated and updated as new information becomes available. This resembles incremental view maintenance, where systems propagate changes to derived representations without recomputing them from scratch. However, unlike traditional views, semantic context is generated from heterogeneous sources—including schemas, documentation, historical queries, and agent trajectories—and changes may propagate through implicit semantic dependencies rather than explicit relational dependencies. At the other extreme, increasingly capable long-context models may make holistic regeneration an attractive alternative. Rather than maintaining complex dependency structures, a system could periodically regenerate the context layer directly from the underlying environment. The appropriate granularity of maintenance introduces a rich design space: fine-grained incremental updates may improve scalability and reduce token costs for large context layers, while holistic regeneration may benefit smaller environments where end-to-end models can reason over the complete context. Understanding this tradeoff—and developing workload-aware strategies for semantic maintenance—represents an important research opportunity.

\subsection{Context Construction and Physical Design}
\label{sec:agenda-construction}
Crucially, the semantic context layer allows redundant online query processing to shift to an offline preprocessing stage that can be amortized. Rather than requiring agents to repeatedly explore schemas, inspect tables, and rediscover organizational knowledge, the agent system constructs a reusable semantic representation, such as natural-language Markdown files. Several existing works demonstrate the promise of performing an offline processing step~\cite{lin2025sleeptime,zhang2025ace,agrawal2025gepa,agentsm2026} to improve the agent's efficiency on downstream queries.
However, building the semantic context layer for general data agents will require processing a substantial data environment and a large volume of historical traces, making scalability and efficiency core problems for the offline build.
An open question remains as to how the community should design techniques for constructing context layers at scale while providing efficient build times, low memory footprints, and fast access patterns. Excitingly, many of these systems challenges are analogous to traditional indexing problems, which the database community is well-equipped to explore.

\heading{Data Structure.}
One central open question is: what data structure should represent semantic context?
Unlike a B-tree or hash index, a semantic context structure must encode heterogeneous information—including schemas, documentation, historical queries, execution traces, and organizational knowledge—while supporting efficient construction, updates, and retrieval.
Existing works have already proposed several candidates, ranging from free-form text files to structured knowledge graphs and vector indexes~\cite{zhang2025ace,edge2024graphrag,lewis_retrieval-augmented_2021}. These techniques present different tradeoffs in build time, memory overhead, update complexity, retrieval efficiency, and freshness. Future research is needed to understand which representations are best suited for different workloads. Much like classical database systems combine multiple physical structures (e.g., indexes, materialized views, and caches), semantic context may also require workload-specific combinations of representations.

\heading{Physical Design.} The physical organization of semantic context presents new opportunities. During execution, agents are likely to rely on both \emph{query-agnostic context}, such as schema descriptions, business terminology, and organizational documentation that can be reused across many executions, and \emph{query-aware context}, such as intermediate reasoning, execution traces, or task-specific insights generated for a particular workload. These heterogeneous access patterns raise classical physical design questions.
Which context should be stored persistently? Which should be cached or indexed? How should semantic memory be partitioned across fast but expensive representations (e.g., KV cache or local text memory) versus slower but compact long-term storage?

\heading{Compression.}
A long-lived semantic context layer may eventually contain millions of tokens of schemas, documentation, historical query traces, execution logs, and agent-generated reasoning, far exceeding what can be consumed directly by an LLM. 
While recent work has explored \emph{compaction} over KV caches~\cite{eyuboglu2025cartridgeslightweightgeneralpurposelong,zweiger2026fastkvcompactionattention} that serve as the agent's working memory over a session, considerably less attention has been paid to compressing persistent semantic memory. Future systems will require efficient techniques to consolidate accumulated knowledge while preserving the information most useful for downstream tasks. More broadly, semantic context introduces a new lifecycle management problem: as agents continuously generate new observations, systems must determine what should be retained, merged, compressed, or discarded, motivating new garbage collection and compression policies.

\section{Conclusion}
\label{sec:conclusion}
We propose a vision for future database research towards enabling data agents amid the rapid evolution of model capabilities. 
Contextual knowledge and understanding over large, complex data environments remains a substantial challenge for state-of-the art agents, making persistent semantic contexts crucial.
Demonstrating both the promise and system challenges associated with persistent semantic contexts, we outline exciting research opportunities for designing data systems that natively support agents with efficient context data structures, storage and compression techniques,  as well as semantic consistency.

\begin{acks}
This research was supported in part by affiliate members and other supporters of the Stanford DAWN project, including Meta, Google, and VMware, as well as Cisco, SAP, and a Sloan Fellowship. Any opinions, findings, and conclusions or recommendations expressed in this material are those of the authors and do not necessarily reflect the views of the sponsors.
\end{acks}

\bibliographystyle{ACM-Reference-Format}
\bibliography{citations}

\onecolumn
\appendix
\section{Experimental Details}
\subsection{Human-designed Baselines}
\label{app:human-baselines}

\heading{Agentar-inspired SQL pipeline.}
We implement a SQL-only approximation of
Agentar-Scale-SQL~\cite{wang2025agentarscalesqladvancingtexttosqlorchestrated}
for the 60 non-aggregation TAG-Bench queries. The model generates SQL from the
original question and live SQLite schema, then receives one round of execution
feedback to refine its query. Because the complete Agentar inference pipeline is
not public, our baseline uses its published generation prompt but does not
reproduce its unreleased multi-generator, revisor, or selector components.

\begin{lstlisting}[caption={Agentar-inspired SQL-generation prompt.},label={lst:agentar-generation}]
Task Overview:
You are a data science expert. Below, you are provided with a
database schema and a natural language question. Your task is to
understand the schema and generate a valid SQL query to answer the
question.

Database Engine:
SQLite

Database Schema:
{db_schema}
This schema describes the database's structure, including tables,
columns, primary keys, foreign keys, and any relevant relationships
or constraints.

Question:
{question}

Instructions:
- Make sure you only output the information that is asked in the
  question. If the question asks for a specific column, make sure to
  only include that column in the SELECT clause, nothing more.
- The generated query should return all of the information asked in
  the question without any missing or extra information.
- Before generating the final SQL query, please think through the
  steps of how to write the query.

Output Format:
In your answer, please enclose the generated SQL query in a code block:
```sql
-- Your SQL query
```

Take a deep breath and think step by step to find the correct SQL query.
\end{lstlisting}

\begin{lstlisting}[caption={Execution-guided refinement prompt.},label={lst:agentar-refinement}]
You are performing execution-guided iterative SQL refinement.

Review the original task, the SQL previously generated, and its
execution result. Decide whether that result fully answers the
original question. If it does not, revise the SQL. You may use your
own knowledge and semantic reasoning about values visible in the
result. The revised SQL must be standalone and its execution must
return exactly the requested answer, with no missing or extra values.
You do not have access to a reference or gold answer.

ORIGINAL TASK:
{original_prompt}

PREVIOUS SQL:
```sql
{previous_sql}
```

EXECUTION RESULT:
{execution_preview}

Return exactly one revised SQL query in a ```sql code block.
\end{lstlisting}

\heading{DeepEye.}
On DAB, we run DeepEye's open-source Supervisor/workflow agent
end-to-end~\cite{Li_2026}. We add data-source connectors for
DuckDB and MongoDB.

\subsection{Context Generation Baselines}
\label{app:context-baselines}

All context-generation experiments use the same fixed stratified split of DAB:
12 development queries (one from each database environment) and 42 held-out
queries. Context is constructed offline with GPT-5.6 Sol.
After construction, each context artifact is frozen and appended to the first user prompt.

\subsubsection{Self-Curated}
\label{app:self-curated}

\heading{Accuracy and latency contexts.}
For each objective, a single generator call receives all 12 development
trajectories: questions, correctness and validator feedback, agent messages,
ordered commands and outputs, token measurements, and latency measurements. Listings
\ref{lst:self-curated-common}--\ref{lst:self-curated-objectives} show the common
generator prompt and the objective-specific suffixes. The generated instruction
blocks, which are injected verbatim during evaluation, appear in
Listings~\ref{lst:self-curated-accuracy} and~\ref{lst:self-curated-latency}.

\begin{lstlisting}[caption={Common Self-Curated generation prompt.},label={lst:self-curated-common}]
You are designing reusable procedural instructions for a database-
question agent. You are running as GPT-5.6 Sol with high reasoning.

You have access to exactly 12 development trajectories from an earlier
GPT-5.6 Sol/high run in `development_traces.jsonl`.

Each JSON line contains the question, correctness and validator feedback,
latency/token metrics, agent messages, and the ordered command trajectory
with tool outputs. These are the only empirical examples you may use.

Analyze all 12 trajectories, including both successes and failures.
Identify recurring root causes, inefficient behaviors, reliable
strategies, and interactions between accuracy and action count.
Distinguish an accidental one-query fix from a reusable behavioral rule.

First build an internal evidence table for every proposed instruction:
(a) the recurring behavior it addresses, (b) how often that behavior
appears, (c) whether it occurs in correct traces, incorrect traces, or
both, (d) its likely accuracy benefit, and (e) its likely latency/action
cost. Do not emit the table, but use it to reject rules supported by only
one idiosyncratic example unless they prevent a broad class of severe
errors.

Produce a compact instruction block that will be injected verbatim into
the FIRST user prompt of unseen held-out database questions. The evaluator
will not mount a skill and the downstream agent must not spend actions
locating or reading one.

Requirements for `instructions`:
- Write standalone imperative guidance, not an analysis or retrospective.
- Generalize across SQLite, DuckDB, PostgreSQL, MongoDB, structured
  metadata, free text, multiple sources, aggregation, ranking, and exact
  formatting.
- Do not mention DataAgentBench, traces, development/held-out splits,
  query IDs, dataset names, ground-truth answers, or example-specific
  constants.
- Do not tell the downstream agent to read a skill or auxiliary
  instruction file.
- Do not prescribe progress messages.
- Preserve the benchmark rules: use only sanctioned databases, execute
  the computation, and return the requested answer format.
- Prefer conditional decision rules over unconditional extra steps. State
  the observable trigger, the action, and the stopping condition.
- Do not impose a single hard action cap across heterogeneous questions.
  If budgets are useful, use risk-sensitive soft budgets with explicit
  escalation triggers.
- Separate indispensable correctness checks from redundant confirmation.
  Require a check only when it can discriminate between plausible answers.
- Avoid comprehensive database-agent advice the base model likely knows.
  Include only trace-supported behavioral deltas.
- Keep the instruction block between 120 and 450 words. Every instruction
  must justify its prompt-token and behavioral cost.
- End `predicted_tradeoffs` with falsifiable predictions for commands/query,
  output tokens/query, latency/query, and accuracy relative to baseline.

Before answering, verify that you analyzed 12 unique trajectories.
Return only the requested JSON object containing `analysis_summary`,
`instructions`, and `predicted_tradeoffs`.
\end{lstlisting}

\begin{lstlisting}[caption={Objective-specific suffixes for Self-Curated context generation.},label={lst:self-curated-objectives}]
[Accuracy]
Primary objective: maximize accuracy on unseen tasks.

Design the smallest set of high-value guardrails that would have
prevented multiple recurring incorrect answers. Prioritize population/
grain mistakes, semantic interpretation, cross-source key alignment,
sparse-field semantics, ranking/ties, date boundaries, parsing coverage,
and exact output shape only when supported by the traces.

Do not turn every risk into a mandatory action. For each check, specify a
detectable risk trigger and a cheap decisive test. Preserve successful
baseline behavior: avoid replacing model judgment with rigid procedures,
avoid repeated schema reads, and avoid verbose planning. Prefer one
composed computation plus one risk-triggered validation over several
incremental probes. If two computations disagree, prescribe how to select
the interpretation grounded in the question and data rather than merely
running more variants.

Target a measurable accuracy improvement without increasing median
commands, tokens, or latency by more than necessary. Remove any
instruction whose only effect is to encourage generic caution.

[Latency]
Primary objective: minimize end-to-end latency and action count on unseen
tasks while maintaining at least the original agent's accuracy.

Reduce model reasoning time as well as tool time. Remove redundant
exploration, fragile commands, repeated schema reads, unnecessary
narration, alternative computations that cannot change the answer, and
unconditional verification. Favor consolidated catalog/schema reads and
one composed robust computation.

Do not obtain speed by prematurely accepting a plausible result. Replace
the previous style of global hard action caps and near-elimination of
verification with a risk-gated policy:
- stop immediately for low-risk direct lookups once the requested result
  shape is obtained;
- perform one cheap targeted check when semantics, joins, dates, ties,
  parsing, sparse fields, or aggregation grain could plausibly change the
  answer;
- escalate beyond the soft budget only after a failed command, low join/
  parser coverage, contradictory evidence, or an unresolved ambiguity
  that affects the result.

Specify which operations should be combined, which small diagnostics are
worth their latency, and an explicit stopping condition after a decisive
check. Preserve correctness-critical semantic processing. The intended
improvement is fewer recovery and redundant analytical actions, not fewer
data-grounding actions.
\end{lstlisting}

\begin{lstlisting}[caption={Generated Self-Curated Accuracy context.},label={lst:self-curated-accuracy}]
Use only the sanctioned databases, execute the computation, and return
the requested answer rather than an unexecuted query.

Inspect the catalog, schema, and representative records once. Translate
the request into the target population, entity grain, filters, measure,
grouping, ranking, and output cardinality. When an entity can have
repeated rows, versions, labels, or events, deduplicate by its canonical
ID at the requested stage; add a distinct-count/cardinality check only if
duplication could change the result.

When combining sources, first test the proposed keys on a few records and
compute match coverage and join fan-out alongside the main computation.
Do not treat a sample name, display text, or repository example as the
canonical owner when an ID or link table exists. Normalize case,
whitespace, prefixes, or suffixes only when the data demonstrates that
mapping.

If the requested concept is a topic, intent, qualification, or category
with no structured field, do not substitute literal keyword occurrence.
Use the relevant text and knowledge evidence with a coverage-oriented
semantic rule; inspect a small contrasting sample and category prevalence,
then stop when the interpretation is stable. Treat a missing row or null
as "absent" only if the source represents the complete eligible population.

If free-text parsing materially drives the answer, measure parse coverage
and ambiguity. Inspect misses or multiple matches only when they could
affect the winner; expand parsing until coverage is complete or the
remaining misses provably cannot change it. Parse date variants before
filtering and express inclusive calendar windows as a half-open range. For
moving-window or EMA calculations, materialize the intended period
sequence and decide whether missing periods are zero or unknown from the
question's meaning.

For top-k results, inspect the cutoff for ties. Use a requested or
data-defined secondary key; otherwise expose the tie rather than inventing
an arbitrary tie-break.

Prefer one composed computation plus one cheap, winner-discriminating
validation. Escalate only for join loss/fan-out, material parser misses,
cutoff ties, sparse-absence uncertainty, or genuinely competing
interpretations. If variants disagree, select the one grounded in the
wording, schema, and data semantics; do not average, vote, or keep querying
without a decision criterion. Finish with exactly the requested fields,
order, count, precision, and no extra explanation when "only" is requested.
\end{lstlisting}

\begin{lstlisting}[caption={Generated Self-Curated Latency context.},label={lst:self-curated-latency}]
Use only the databases authorized in the supplied catalog/description,
execute the computation against them, and return exactly the requested
fields, ordering, rounding, and format.

Start from the supplied catalog; do not scan the filesystem when
connection details are already present. In one initial command, read the
catalog/description and inspect only the relevant schemas, value samples,
and basic counts across candidate sources. Use `python3` directly for
cross-database orchestration, open analytical files read-only, quote
unusual identifiers, and avoid reserved aliases.

Before computing, fix the semantic contract internally: eligible
population, entity grain and unique key, join keys, date bounds, metric,
grouping, ranking, tie policy, and output shape. Honor explicit key/grain
language: deduplicate or count references by the stated ID before
attributing results to a repository, package, business, or other parent.
Prefer structured fields and relations; do not substitute a free-text
keyword for a category label unless the data establishes that encoding.

Compose filtering, key normalization, deduplication, joining, aggregation,
ranking, and formatting into one robust computation. When prose parsing is
unavoidable, handle observed wording variants and retain matched/unmatched
counts. If joins or parsers affect the answer, perform one cheap coverage
diagnostic and inspect only unmatched records capable of changing the
winner. For dates, rolling statistics, or EMA, check inclusivity,
missing-period treatment, and initialization only when those choices could
alter the result. For top-N results, inspect cutoff ties and never invent
an arbitrary tie-break.

Stop immediately after a low-risk direct lookup produces the requested
shape. Otherwise perform one targeted check comparing the plausible
alternative that could change the answer; stop once it preserves the
winner or resolves the ambiguity. Treat 2-3 tool actions as a soft target
for direct structured lookups and 3-5 for multi-source, parsing, or
statistical tasks. Exceed these only after a failed command, low coverage,
contradictory evidence, or unresolved result-changing ambiguity. Do not
repeat schema reads or computations that cannot change the answer.
\end{lstlisting}

\subsubsection{GEPA}
\label{app:gepa}

We use GEPA version 0.1.1~\cite{agrawal2025gepa} to optimize a persistent
instruction prompt for accuracy. Optimization starts from an empty instruction
and uses the same 12 development tasks as both training and validation data.
Each candidate is evaluated through the same isolated Codex harness used for the no-context baseline. We cap optimization at 60 logical metric evaluations.



\end{document}